\documentclass
{elsart5p}

\usepackage{times}
\usepackage{mathrsfs}

\usepackage{graphicx}

\usepackage{amsmath}
\usepackage{amssymb}

\renewcommand{\vec}[1]{\boldsymbol{#1}}      
\renewcommand{\geq}{\geqslant}

\allowdisplaybreaks

\begin{document}
\parindent 1em

\begin{frontmatter}
\vspace{-50pt}


\title{Model of hopping dc conductivity via nearest neighbor boron atoms in~moderately compensated diamond crystals}


\author[BSU]{N.A. Poklonski\corauthref{cor}},
\corauth[cor]{Corresponding author.}
\ead{poklonski@bsu.by}
\author[BSU]{S.A. Vyrko},
\author[PTI]{A.G. Zabrodskii}

\address[BSU]{Belarusian State University, Minsk 220030, Belarus}
\address[PTI]{Ioffe Physicotechnical Institute RAS, St. Petersburg 194021, Russia}

\begin{abstract}
Expressions for dependences of the pre-exponential factor $\sigma_3$ and the thermal activation energy $\varepsilon_3$ of hopping electric conductivity of holes via boron atoms on the boron atom concentration $N$ and the compensation ratio $K$ are obtained in the quasiclassical approximation. It is assumed that the acceptors (boron atoms) in charge states ($0$) and ($-1$) and the donors that compensate them in the charge state ($+1$) form a nonstoichiometric simple cubic lattice with translational period $R_\text{h} = [(1 + K)N]^{-1/3}$ within the crystalline matrix. A hopping event occurs only over the distance $R_\text{h}$ at a thermally activated accidental coincidence of the acceptor levels in charge states ($0$) and ($-1$). Donors block the fraction $K/(1 - K)$ of impurity lattice sites. The hole hopping conductivity is averaged over all possible orientations of the lattice with respect to the external electric field direction. It is supposed that an acceptor band is formed by Gaussian fluctuations of the potential energy of boron atoms in charge state ($-1$) due to Coulomb interaction only between the ions at distance $R_\text{h}$. The shift of the acceptor band towards the top of the valence band with increasing $N$ due to screening (in the Debye--H\"uckel approximation) of the impurity ions by holes hopping via acceptor states was taken into account. The calculated values of $\sigma_3(N)$ and $\varepsilon_3(N)$ for $K \approx 0.25$ agree well with known experimental data at the insulator side of the insulator--metal phase transition. The calculation is carried out at a temperature two times lower than the transition temperature from hole transport in $v$-band of diamond to hopping conductance via boron atoms.

\vspace{-12pt}
\end{abstract}

\end{frontmatter}

\section{Introduction}

Following the observation of helium temperature superconductivity in heavily boron doped diamond, numerous studies of this material were performed (see, e.g., reviews~\cite{Bustarret08, Mares08, Gajewski09}). The possible use of intermediately boron doped diamond in semiconductor applications~\cite{Wort08} justifies studies of its conductivity at room temperature. With temperature lowering from the room temperature to the temperature of liquid nitrogen a conduction of holes in $v$-band (propagating regime \cite{Datta80}) changes into hopping conduction of holes via boron atoms. The dc hopping conduction in boron-doped diamond is observed in the dark at significantly higher temperatures and concentrations of boron than in silicon and germanium crystals doped with the same acceptor impurity at comparable compensation ratios. Because of the progress in the synthesis technology of high-quality homoepitaxial crystalline diamond films with controllable boron doping, reliable experimental data on the hopping conductivity $\sigma_\text{h}$ via boron atoms comparable with computational models was eventually obtained~\cite{Borst96, Malta95, Visser92}. However, a satisfactory quantitative description of the hopping transport of holes in diamond crystals is still lacking. 


In this paper we limit our consideration to the hole hopping regime via nearest neighbor boron atoms (NNH regime). In this case, the dc hopping conductivity is (see, e.g., Ref.~\cite{Shklovskii84})
\begin{equation}\label{eq:01}
	\sigma_\text{h} = \sigma_3 \exp\biggl(-\frac{\varepsilon_3}{k_\text{B}T}\biggr),
\end{equation}
where $\sigma_3 \equiv 1/\rho_3$ is the pre-exponential factor, $\varepsilon_3/k_\text{B}T$ is the ratio of the activation energy of hole transport via impurity atoms to the thermal energy. 

When describing small polaron hopping over the lattice sites in an ionic crystal, Holstein (see reprint~\cite{Holstein00}) introduced the concept of a ``coincidence event'' for polaron potential wells. 
In the model~\cite{Holstein00} (see also Refs.~\cite{Heikes61, Nagel's82}), a polaron hop is assumed to occur when the energies of the initial occupied state and the final vacant state coincide. 

In Ref.~\cite{Burin89}, a model of fluctuation-induced ``alignment'' of the energy levels of localized states (of impurity atoms) due to electron-electron interaction was proposed to describe the dc conduction of doped semiconductors. It was assumed that temporal fluctuations of the energy of localized states are caused by hopping diffusion of electrons via these states. Another model (the variable range hopping (VRH) conduction model~\cite{Kozub00}) has been proposed, assuming that electron (or hole) hops occur via resonance tunneling between atoms of majority impurity. Energy levels of two impurity atoms enter into resonance due to Coulomb potential fluctuations induced by stochastic changes in the occupation state of other doping impurity atoms.\footnote{The model of the fluctuation-induced preparation of a barrier through which an atom (or even a molecule) can tunnel made it possible to explain the main characteristics of solid-phase cryochemical reactions (see, e.g.,~\cite{Gol'danskij86}).} However, in studies~\cite{Burin89, Kozub00} numerical calculations of the experimentally observed quantities were not carried out.

In Refs.~\cite{Poklonskij00FTT432, Poklonskij01FTT2126}, the dc hopping conductivity and thermoelectric power in the NNH regime were described for neutron transmutation doped \emph{p} type germanium. It was supposed that the doping impurity (acceptors) and a compensating impurity (donors) form a common ``impurity lattice'' in the crystalline matrix. It was assumed that hole hopping occurs only during thermal-induced alignment (coincidence) of energy levels of the acceptors in charge states ($0$) and ($-1$). Donors, which are all in charge state ($+1$), block some sites of the impurity lattice. 
At the instant of energy level coincidence (up to a broadening of the acceptor energy levels due to the finite time of hole localization on the acceptor) of a neutral and a negatively charged acceptor, a ``resonant'' two-site cluster is formed: the hole on acceptor 1 becomes bound to a negatively charged acceptor 2 and belongs simultaneously to these two acceptors. After some time, the resonance conditions are no longer satisfied and the hole can become localized on acceptor 2 or remain on acceptor 1. After that, the acceptors 1 and 2 can again form a resonant cluster or form resonant clusters (acceptor pairs) with other acceptors.

The aim of our work is to determine the dependences of $\sigma_3$ and $\varepsilon_3$ in Eq. \eqref{eq:01} on the concentration of boron atoms in moderately compensated, intermediately doped diamond crystals. For this purpose the model from Ref.~\cite{Poklonskij00FTT432} is developed.

\section{Statistics of hydrogen-like impurities}

Let us consider a \emph{p} type uniform crystalline semiconductor with acceptor concentration $N = N_0 + N_{-1}$, where $N_0$ and $N_{-1}$ are the concentrations of acceptors in charge states ($0$) and ($-1$), respectively. There are also donors, all in charge state ($+1$), with concentration $KN$, where $0 < K < 1$ is a compensation ratio of acceptors by donors.\footnote{We consider the boron concentrations $N$ much less than the concentration at which in diamond the transition from insulator state to metallic one (Mott transition) $N_\text{M} \approx 2{\cdot}10^{20}$ cm$^{-3}$ is observed~\cite{Borst96}. The calculations of $N_\text{M}$ dependence on compensation ratio $K$ were carried out in the work~\cite{Poklonski09pssb158}.} The electrical neutrality condition has the form
\begin{equation}\label{eq:02}
	KN = N_{-1}.
\end{equation}
According to Eq. \eqref{eq:02}, the probability that a randomly chosen acceptor is in charge state ($-1$) or ($0$) is equal to $K$ or $(1 - K)$, respectively. Thus, the average (over the crystal) fraction of acceptors in charge state ($-1$) is~\cite{Poklonski09pssb158, Poklonskij99FTP415, Poklonski83}
\begin{equation}\label{eq:03}
	\frac{N_{-1}}{N} = \int_{-\infty}^{+\infty}\mathcal{G}f_{-1}\,\mathrm{d}(E - \overline{E}) = K,
\end{equation}
where $\mathcal{G}$ is the distribution of acceptor energy levels $E$ with respect to the average value $\overline{E}$ when $f_{-1}$ is the probability that acceptor with energy level $E$ is ionized. Further, we assume that $\mathcal{G}$ is a Gaussian distribution:
\begin{equation}\label{eq:04}
	\mathcal{G} = \frac{1}{\sqrt{2\pi}W} \exp\biggl[\frac{-(E - \overline{E})^2}{2W^2}\biggr],
\end{equation}
where $W$ is the effective width of the acceptor band. According to Refs.~\cite{Poklonski09pssb158, Blakemore02} the probability that an acceptor with energy level $E > 0$ above the top of the $v$-band ($E_v = 0$) of an undoped crystal is ionized can be written as
\begin{align}\label{eq:05}
	f_{-1} = 1 - f_0 = \biggl[1 + \beta\exp\biggl(\frac{E + E_\text{F}}{k_\text{B}T}\biggr)\biggr]^{-1} \notag\\ 
	\equiv [1 + \exp(u + \zeta)]^{-1},
\end{align}
here $\beta = 6$ is the degeneracy factor of the energy level of a boron atom in diamond, $E_\text{F}$ is the Fermi level relative to the top of the $v$-band ($E_\text{F} < 0$ in the band gap of diamond), $k_\text{B}T$ is the thermal energy, $u = (E - \overline{E})/k_\text{B}T$ and $\zeta - \ln\beta = (E_\text{F} + \overline{E})/k_\text{B}T$ are dimensionless acceptor energy level and Fermi level relative to the center $\overline{E}$ of the acceptor band.

We assume that the doping impurity (hydrogen-like acceptors) and the compensating impurity (hydrogen-like donors) form a common nonstoichiometric simple cubic lattice within the crystalline matrix (cf.~\cite{Mycielski61}). The translational period of this lattice is $R_\text{h} = [(1 + K)N]^{-1/3}$, where $(1 + K)N$ is the total concentration of impurities. In further consideration we suppose that $R_\text{h}$ is the length of hole hop between acceptors in charge states ($0$) and ($-1$) in the impurity lattice.

The acceptor band width, taking into account the Coulomb interaction of the ionized acceptor with the ions in the first coordination sphere of the impurity lattice, is equal to~\cite{Poklonskij00FTT432, Poklonskij01FTT2126}
\begin{equation}\label{eq:06}
	W = \biggl(\sum_{i=1}^6\mathcal{P}_iU_i^2\biggr)^{\!1/2} = \frac{e^2}{4\pi\varepsilon R_\text{h}} \biggl(\frac{12K}{1 + K}\biggr)^{\!1/2},
\end{equation}
where $\mathcal{P}_i = 2K/(1 + K) = 2K\Xi$ is the probability that any of 6 sites of a simple cubic impurity lattice (in the first coordination sphere) near the $i$-th ion is occupied by an ionized acceptor or donor, $\Xi = 1/(1 + K)$ is the correlation factor, i.e., the fraction of majority impurity at impurity lattice sites, $|U_i| = e^2/(4\pi\varepsilon R_\text{h})$ is the magnitude of the Coulomb interaction energy between the $i$-th ion and an ion in the impurity lattice at distance $R_\text{h}$ from it, $e$ is the modulus of the electron charge, $\varepsilon = 5.7\varepsilon_0$ is the static permittivity of diamond, $\varepsilon_0$ is the electric constant. For derivation of Eq. \eqref{eq:06} it is taken into account that the Coulomb interaction energy of the ions in the first coordination sphere averaged over the impurity lattice is equal to zero: $\sum_{i=1}^6\mathcal{P}_iU_i = 0$.

According to Eq. \eqref{eq:06}, we assume that the acceptor band is a ``classical'' one, i.e., the spread of the energy levels of the boron atoms is much greater than the quantum-mechanical broadening of these levels due to the finite time of hole localization on the acceptor.

The position of the acceptor band center $\overline{E}$ relative to the top of the $v$-band ($E_v = 0$), according to Refs.~\cite{Poklonski09pssb158, Poklonskij99FTP415}, is
\begin{equation}\label{eq:08}
	\overline{E} = I - \frac{3e^2}{16\pi\varepsilon(\Lambda_\text{h} + R_\text{h})},
\end{equation}
where $I = 370$ meV is the energy level of a single acceptor (boron atom in diamond), $\Lambda_\text{h}$ is a screening radius of the Coulomb field of the ion, $R_\text{h} = [(1 + K)N]^{-1/3}$ is the minimal possible distance between ions in the impurity lattice. In the Debye--H\"uckel approximation, according to Refs.~\cite{Poklonski09pssb158, Poklonskij99FTP415, Poklonski83}, we find
\begin{equation}\label{eq:09}
	\Lambda_\text{h}^{-2} 
	= \frac{e^2K(1 - K)N}{\varepsilon k_\text{B}T\xi_\text{h}},
\end{equation}
where $\xi_\text{h} \geq 1$.
In accordance with Ref.~\cite{Poklonski83}, the reciprocal value of the quantity $\xi_\text{h}$ in the modified (generalized) Einstein relation for hopping migration of holes via acceptors is determined by expression:
\begin{equation}\label{eq:07}
	\frac{1}{\xi_\text{h}} = \frac{\mathrm{M}_\text{h}k_\text{B}T}{eD_\text{h}} = \frac{1}{K(1 - K)} \int_{-\infty}^{+\infty}\mathcal{G}f_0f_{-1}\,\mathrm{d}(E - \overline{E}),
\end{equation}
where $D_\text{h}/\mathrm{M}_\text{h}$ is the ratio of the diffusion coefficient of hopping holes to their mobility.

For the narrow acceptor band ($W \ll k_\text{B}T$), taking into account Eqs. \eqref{eq:04}--\eqref{eq:06} it follows from Eq. \eqref{eq:07} that the Einstein relation is fulfilled in the classical form (i.e., $\xi_\text{h} = 1$). In this case, we find from Eq. \eqref{eq:02} the quantity $\zeta = \ln\beta + (E_\text{F} + \overline{E})/k_\text{B}T \approx -\ln[K/(1 - K)]$ taking into account Eqs. \eqref{eq:03}--\eqref{eq:05}.

For the wide acceptor band ($W \gg k_\text{B}T$), according to Ref.~\cite{Poklonski83}, we obtain from Eq. \eqref{eq:07}:
\[
	\xi_\text{h} \approx K(1 - K)\gamma\sqrt{2\pi}\exp(\zeta^2/2\gamma^2),
\]
where $\gamma = W/k_\text{B}T \gg 1$. In this case, Eq. \eqref{eq:02} takes the form: $2K \approx 1 - \mathrm{erf}(\zeta/\sqrt{2}\gamma)$.

\section{Hopping current and conductivity}

Let us consider the range of $N$, $K$ and $T$ values at which the NNH regime is only realized, and $\varepsilon_3$ weakly decreases upon a reduction in temperature. 

According to Refs.~\cite{Poklonskij00FTT432, Poklonskij01FTT2126, Poklonskij00FTT218, Poklonskij08FTP1420}, a stationary hopping current density $J_\text{h}$ of holes over acceptors in a sample subjected to an external electric field of the strength $\mathscr{E} = -\mathrm{d}\varphi/\mathrm{d}x$ directed along the $x$ axis has the form:
\begin{equation}\label{eq:10}
	J_\text{h} = eN_\text{h} \biggl(\mathrm{M}_\text{h}\mathscr{E} - D_\text{h}\frac{\mathrm{d}}{\mathrm{d}x} \ln\frac{N_0}{N_{-1}}\biggr) = \sigma_{\text{h}x}\mathscr{E} - eD_\text{h}\frac{\mathrm{d}N_0}{\mathrm{d}x},
\end{equation}
where $N_\text{h} = N_0N_{-1}/N$ is the effective concentration of holes which hop between acceptors in charge states ($0$) and ($-1$), $\mathrm{M}_\text{h} = \Xi R_\text{h}^2\,\mathrm{d}\Gamma/\mathrm{d}\varphi > 0$ is the hopping mobility of holes, $\Xi = 1/(1 + K)$ is the fraction of acceptors at impurity lattice sites, $R_\text{h} = [(1 + K)N]^{-1/3}$ is the length of the hole hop, $(-\mathrm{d}\Gamma/\mathrm{d}x)R_\text{h}$ is the difference of the average hole hopping frequency in the direction along and against the external electric field, $\varphi(x)$ is the electric potential, $D_\text{h} = \Xi R_\text{h}^2\Gamma_\text{h}/6$ is the diffusion coefficient, $\Gamma_\text{h}/6$ is the average hole hopping frequency along one of the six directions (along the edges) of the impurity lattice for zero external field. 
The hopping conductivity $\sigma_{\text{h}x}$ of holes in the impurity lattice for external field $\mathscr{E}$ orientation along the edge of the unit cell of the cubic impurity lattice is given by Eq. \eqref{eq:10}, taking into account Eq. \eqref{eq:02}, in the form
\begin{equation}\label{eq:11}
	\sigma_{\text{h}x} = eN_\text{h}\mathrm{M}_\text{h} = eK(1 - K)N\mathrm{M}_\text{h}.
\end{equation}
Let us find from Eq. \eqref{eq:07}, using $D_\text{h} = \Xi R_\text{h}^2\Gamma_\text{h}/6$, the hole mobility $\mathrm{M}_\text{h}$ in terms of the equilibrium frequency $\Gamma_\text{h}$ of their hopping via acceptors
\[
	\mathrm{M}_\text{h} = \frac{eD_\text{h}}{\xi_\text{h}k_\text{B}T} = \frac{e\Xi R_\text{h}^2\Gamma_\text{h}}{6\xi_\text{h}k_\text{B}T}.
\]
Thus, from Eq. \eqref{eq:11} we obtain the hopping conductivity due to hole hopping with hop length $R_\text{h}$ along the external electric field (along the $x$ axis)
\begin{equation}\label{eq:12}
	\sigma_{\text{h}x} = \frac{e^2\Xi K(1 - K)NR_\text{h}^2\Gamma_\text{h}}{6\xi_\text{h}k_\text{B}T}.
\end{equation}

Let us take into account all possible orientations of the cubic impurity lattice with respect to the direction of the external electric field $\mathscr{E}$. The space of possible orientations is a semi-sphere with a normalized element of the surface $(1/2\pi)\sin\theta\,\mathrm{d}\varphi\,\mathrm{d}\theta$. All edges of the impurity lattice unit cells, each of length $R_\text{h}$, directed at angle $\theta$ to the field strength $\mathscr{E}$ make a contribution $\sigma_{\text{h}x}\cos\theta$ to the conductivity, where $\sigma_{\text{h}x}$ is given by Eq. \eqref{eq:12}. As a result, taking into account Eq. \eqref{eq:01}, we find:
\begin{align}\label{eq:13}
	\sigma_\text{h} = \frac{\sigma_{\text{h}x}}{2\pi} \int_0^{2\pi}\mathrm{d}\varphi \int_0^{\pi/2}\cos\theta\sin\theta\,\mathrm{d}\theta = \frac{\sigma_{\text{h}x}}{2} \notag\\ 
	= \sigma_3\exp\biggl(-\frac{\varepsilon_3}{k_\text{B}T}\biggr).
\end{align}

To obtain the dependence of hopping conductivity $\sigma_\text{h}$ on the values of $N$, $K$ and $T$ using Eqs. \eqref{eq:13} and \eqref{eq:12} one needs to find the frequency of hole hopping $\Gamma_\text{h}$ in the impurity lattice.

For the not very small and not very large ratios of acceptor compensation by donors (tentatively for $0.1 < K < 0.9$) the correlation between the position of an acceptor in the impurity lattice with translational period $R_\text{h}$ and the acceptor energy level $E$ may be neglected. Following Ref.~\cite{Poklonskij00FTT432}, we suppose that every hole hop of length $R_\text{h}$ between acceptors 1 and 2 in charge states ($0$) and ($-1$) occurs only when their energy levels ($E_1 = \overline{E} + u_1k_\text{B}T$ and $E_2 = \overline{E} + u_2k_\text{B}T$) accidentally coincide. The number of hole transitions between acceptors (boron atoms) per energy level coincidence event is equal to the integer part of the ratio of duration $t_k(u)$ of an event of coincidence of levels ($u_1 = u_2 = u = (E_\tau - \overline{E})/k_\text{B}T)$ to the time $\tau(u)$ of a tunneling event. We assume also that for the time interval $t$ the total duration of all events of level coincidence is equal to $t_\mathrm{c}(u) = \sum_kt_k(u)$. We approximate the conditional probability that exactly $j$ transitions of a hole occur at the coincidence of levels of two nearest neighbor acceptors by the Poisson distribution~\cite{Koks66, Uittl82}:
\begin{equation}\label{eq:14}
	P\{j|u\} = \frac{[t_\mathrm{c}(u)/\tau(u)]^j}{j!} \exp\biggl[-\frac{t_\mathrm{c}(u)}{\tau(u)}\biggr],
\end{equation}
where $t_\mathrm{c}(u)/\tau(u) = \sum_{j=0}^\infty jP\{j|u\}$ is the average number of hole transitions between nearest acceptors (boron atoms), $\tau(u)$ is the duration of a tunneling transition of a hole from the neutral acceptor to the ionized one, $j = 0, 1, 2, 3, \ldots$.

Thus, the frequency of hole hopping between two acceptors at accidental alignment of their energy levels $E_\tau = \overline{E} + uk_\text{B}T$ (an average number of hole transitions for the time~$t$) is
\begin{equation}\label{eq:15}
	\Gamma(u) = \frac{1}{t} \sum_{j=0}^\infty jP\{j|u\} = \frac{t_\mathrm{c}(u)}{t\tau(u)}.
\end{equation}

From the theory of Markov chains~\cite{Koks66, Uittl82}, it follows that, if the hole transitions between two acceptors are observed over a long time interval ($t \gg \tau(u)$), then the fraction of time spent by the acceptors in  one of two possible states (when their energy levels are coincident or noncoincident) is approximately equal to the stationary probability of the acceptors being in these states. Thus, the ratio $t_\mathrm{c}(u)/t \ll 1$ is approximately equal to the probability that the energy levels of two nearest neighbor acceptors in charge states ($0$) and ($-1$) are aligned (energy values belong to interval $(u, u + \mathrm{d}u)$):
\begin{equation}\label{eq:16}
	\frac{t_\mathrm{c}(u)}{t}\,\mathrm{d}u = P(u)\,\mathrm{d}u = \mathcal{G}(u)\frac{f_0(u)f_{-1}(u)}{K(1 - K)}\,\mathrm{d}u,
\end{equation}
where the product $f_0(u)f_{-1}(u)$ is obtained using Eq. \eqref{eq:05} and the distribution density $\mathcal{G}(u)$ is given by Eq. \eqref{eq:04} at the substitution of $E - \overline{E}$ by $uk_\text{B}T$. The quantity $P(u)\,\mathrm{d}u$ in formula \eqref{eq:16} gives the conditional probability that energy levels of a randomly chosen pair of acceptors in the charge states ($0$) and ($-1$) belong to the interval $(u, u + \mathrm{d}u)$ (see Appendix).

Further, we take into account that the energy level $E_\tau(u) = \overline{E} + uk_\text{B}T > 0$ above the top of the $v$-band is associated with a radius $a_\tau(u) = e^2/(8\pi\varepsilon E_\tau)$ of hole localization on the acceptor with ionization energy $E_\tau$. The Bohr radius for the center ($u = 0$) of the acceptor band is $a_\tau(0) = e^2/(8\pi\varepsilon\overline{E})$.

Within the framework of the theory of the hydrogen molecular ion (H$_2^+$) \cite{Davydov73} a duration of hole tunneling between two acceptors at the distance $R_\text{h}$ when their energy levels coincide ($u_1 = u_2 = u$) can be estimated as~\cite{Bljumenfel'd67, Kejn73}:
\begin{equation}\label{eq:17}
	\tau(u) = \frac{\pi\hbar}{\delta E_\tau(u)},
\end{equation}
where $\hbar$ is the reduced Planck constant, $\delta E_\tau(u)$ is the broadening (splitting) of the energy levels $E_\tau(u) = \overline{E} + uk_\text{B}T = e^2/[8\pi\varepsilon a_\tau(u)]$ of the acceptors when the hole tunnels between them:
\begin{align}\label{eq:18}
	&\delta E_\tau(u) = 4E_\tau(u) \notag\\ 
	&\times\frac{\rho(1 + \rho)\exp(-\rho) - [1 - (1 + \rho)\exp(-2\rho)]S}{\rho(1 - S^2)},
\end{align}
\[
	\rho(u) = R_\text{h}/a_\tau(u),\quad
	S(u) = [1 + \rho + (\rho^2/3)]\exp(-\rho).
\]

Let us average $\Gamma(u)$ over the distribution of energy levels which form the acceptor band with effective width $W$ (see Eq.~\eqref{eq:06}). Taking into account Eqs. \eqref{eq:15} and \eqref{eq:16}, the average hole hopping frequency $\Gamma_\text{h}$ between two acceptors at the distance $R_\text{h}$ in the impurity lattice can be written in the form:
\begin{equation}\label{eq:19}
	\Gamma_\text{h} = \int_{-\infty}^{+\infty} \Gamma(u)\,\mathrm{d}u = \int_{-\infty}^{+\infty} \frac{P(u)}{\tau(u)}\,\mathrm{d}u.
\end{equation}

The duration of hole tunneling $\tau(u)$ according to Eqs. \eqref{eq:17} and \eqref{eq:18} monotonically increases when the tunneling level $E_\tau$ moves deeper into the band gap. The function $P(u) \propto \mathcal{G}(u)f_0(u)f_{-1}(u)$ has its sharp maximum at value $u = u_\mathrm{m}$ which satisfies the equation
\begin{equation}\label{eq:20}
	u_\mathrm{m} + \gamma^2\tanh[(u_\mathrm{m} + \zeta)/2] = 0,
\end{equation}
where $\gamma = W/k_\text{B}T$, $\zeta = \ln\beta + (E_\text{F} + \overline{E})/k_\text{B}T$. This allows us to take $\tau(u)$ out of the integral in Eq. \eqref{eq:19} at $u = u_\mathrm{m}$ denoting $\tau(u_\mathrm{m}) = \tau_3$. It follows from Eq. \eqref{eq:20} that if temperature $T \to 0$ then $u_\mathrm{m} \to -\zeta$, and if acceptor band width $W \to 0$ then $u_\mathrm{m} \to 0$. Thus, taking into account Eq. \eqref{eq:07}, the average (over the impurity lattice) frequency of hole hopping is
\begin{align}\label{eq:21}
	\Gamma_\text{h} \approx \frac{1}{\tau_3K(1 - K)} \int_{-\infty}^{+\infty} \mathcal{G}f_0f_{-1}\,\mathrm{d}u = \frac{1}{\tau_3\xi_\text{h}} \notag\\
	\equiv \Gamma_3\exp\biggl(-\frac{\varepsilon_3}{k_\text{B}T}\biggr),
\end{align}
where $\Gamma_3 = 1/\tau(u_\mathrm{m}) \equiv 1/\tau_3$ is the frequency of hole tunneling between acceptors in charge states ($0$) and ($-1$) with energy levels $E_\tau(u_\mathrm{m}) = \overline{E} + u_\mathrm{m}k_\text{B}T$. The electrical neutrality condition \eqref{eq:02} can be solved for the dimensionless Fermi level $\zeta - \ln\beta = (E_\text{F} + \overline{E})/k_\text{B}T$. Calculating the position of the acceptor band center $\overline{E}$ using Eq. \eqref{eq:08}, we find the Fermi level value $E_\text{F} = k_\text{B}T(\zeta - \ln\beta) - \overline{E} < 0$ relative to the top of the $v$-band. The value $u_\mathrm{m}$ is calculated from Eq. \eqref{eq:20}.

We find the pre-exponential factor in the temperature dependence \eqref{eq:01} from expression \eqref{eq:13} taking into account Eqs. \eqref{eq:12} and \eqref{eq:21}:
\begin{equation}\label{eq:22}
	\sigma_3 = \frac{e^2\Xi K(1 - K)NR_\text{h}^2\Gamma_3}{12\xi_\text{h}k_\text{B}T} = \frac{e^2K(1 - K)N^{1/3}\Gamma_3}{12(1 + K)^{5/3}\xi_\text{h}k_\text{B}T},
\end{equation}
where the duration of hole tunneling $1/\Gamma_3 = \tau_3 = \tau(u_\mathrm{m})$ is determined using Eq. \eqref{eq:17} for $u = u_\mathrm{m}$ from Eq. \eqref{eq:20}, and the factor $\xi_\text{h} \geq 1$ is given by formula \eqref{eq:07}.

The activation energy of hopping conductivity follows from Eq. \eqref{eq:21} using Eqs. \eqref{eq:12} and \eqref{eq:13} in the form:
\begin{equation}\label{eq:23}
	\varepsilon_3 = -k_\text{B}T\ln(\Gamma_\text{h}/\Gamma_3) = k_\text{B}T\ln\xi_\text{h}.
\end{equation}
It follows from Eq. \eqref{eq:23} using Eq. \eqref{eq:07} that $\varepsilon_3 \to 0$ at the acceptor band width $W \to 0$ .

According to Eq. \eqref{eq:23} and taking into account Eq. \eqref{eq:07}, the activation energy $\varepsilon_3$ decreases sublinearly upon lowering the temperature $T$, because $\xi_\text{h}$ increases in this case. Let us define the characteristic temperature value $T_\text{h}$ at which $\varepsilon_3$ is measured. For some temperature $T_\text{j}$ the conductivity $\sigma_\text{p}$ of holes in the $v$-band is equal to the hopping conductivity $\sigma_\text{h}$ of holes via boron atoms. The dependence of the temperature $T_\text{j}$ on the concentration $N$ of boron atoms is determined in the same way as in Ref.~\cite{Poklonski09pssb158}, i.e., from the dependences of the logarithm of the total electric conductivity $\sigma = \sigma_\text{p} + \sigma_\text{h}$ on the reciprocal temperature $1/T$ for different $N$ and $0.05 < K < 0.5$, using experimental data~\cite{Borst96, Malta95, Visser92}.\footnote{In the works~\cite{Borst96, Malta95} the total concentration $N = N_0 + N_{-1}$ of boron atoms in diamond was determined from the secondary ion mass spectroscopy measurements. In the work~\cite{Visser92} the concentration $N_0$ of electrically neutral boron atoms was measured from the one-phonon band of IR absorption at $T = 300$ K. Then we estimated for $K \approx 0.25$ the boron concentration to be $N = N_0/(1 - K)$.} We derived the numerical dependence of $T_\text{j}$ on $N$ in the form 
\begin{equation}\label{eq:24}
	T_\text{j} \approx 2N^{0.11},
\end{equation}
where $[T_\text{j}] ={}$K, $[N] ={}$cm$^{-3}$. 

The average temperature $T_\text{h}$, at which the values of $\sigma_3$ and $\varepsilon_3$ were measured in the experiments~\cite{Borst96, Malta95, Visser92}, can be estimated as $T_\text{h} = T_\text{j}/2$, i.e., it is assumed to be equal to the average value in the interval from absolute zero to $T_\text{j}$. (It is likely that in the temperature interval from $T_\text{j}$ to $T_\text{h}$ the NNH regime of hole hopping via boron atoms dominates, and for the temperature below $T_\text{h}$ the VRH regime occurs.) 

\begin{figure}
\noindent\hfil\includegraphics{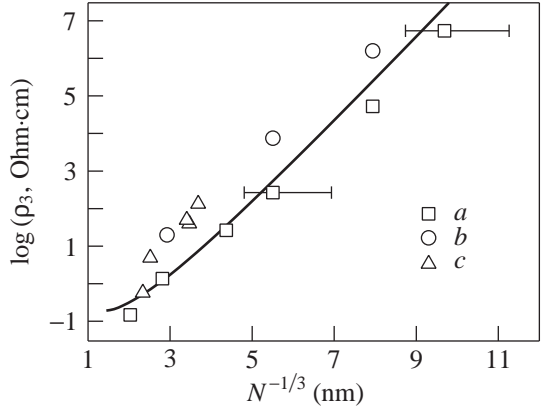} 
\caption{Dependence of the pre-exponential factor $\rho_3 = 1/\sigma_3$ on the concentration $N$ of boron atoms in diamond. Solid line is our calculation using Eq. \eqref{eq:22} for the compensation ratio $K = 0.25$ at the temperature $T_\text{h} = T_\text{j}/2$ defined by Eq. \eqref{eq:24}. Experimental data of papers~\cite{Borst96}, \cite{Malta95}, \cite{Visser92} are denoted by $a$, $b$, $c$, respectively}\label{fig:01}
\end{figure}

In Fig. 1, the dependence of the reciprocal value of the pre-exponential factor \eqref{eq:22} for hopping resistivity ($\rho_3 = 1/\sigma_3$) on the concentration $N$ of boron atoms in diamond (for the compensation ratio $K = 0.25$ and temperature $T_\text{h} = T_\text{j}/2$) is shown by the solid line. When we change $N^{-1/3}$ from 2 to 11 nm the temperature $T_\text{h}$ changes, according to Eq. \eqref{eq:24}, between 160 and 90 K. In this case the ratio $W/k_\text{B}T_\text{h}$ changes approximately from 15 to 5. Thus, according to Eq. \eqref{eq:20} we obtain $u_\text{m} \approx -\zeta$. 

It should be noted that the calculation according to Eq. \eqref{eq:22} also agrees well (see Ref.~\cite{Poklonskij00FTT432}) with the data for $\rho_3$ in \emph{p}~Ge:Ga at $K = 0.35$ and $T_\text{h} = T_\text{j}/2$, where according to Ref.~\cite{Poklonskij99FTP415}, the temperature $T_\text{j} = 5.3{\cdot}10^4N^{0.27}$, $[T_\text{j}] ={}$K, $[N] ={}$cm$^{-3}$. However, in this case, the acceptor energy level of the single gallium atom in germanium is $I = 11.3$ meV, relative permittivity is $\varepsilon/\varepsilon_0 = 15.4$, and the value $N^{-1/3}$ changes approximately from 30 to 140 nm.

In Fig. 2, the dependence of the activation energy $\varepsilon_3$ of hopping conductivity on the concentration $N$ of boron atoms in diamond for compensation ratio $K = 0.25$ is shown by the solid line. The calculation was performed using Eq. \eqref{eq:23} taking into account Eq. \eqref{eq:07} at the temperature $T_\text{h} = T_\text{j}/2$. In Fig. 2, the calculation of $\varepsilon_3 \approx 0.7e^2N^{1/3}/(4\pi\varepsilon)$ according to the model~\cite{Shklovskii84}, for $K \approx 0.25$ and relative permittivity of diamond $\varepsilon/\varepsilon_0 = 5.7$, is shown by the dashed line. It can be seen that the calculation according to Eq. \eqref{eq:23} agrees well with the experimental data, while the calculation using the model~\cite{Shklovskii84} gives overestimated values of the activation energy $\varepsilon_3$.

Note that, for all covalent semiconductors with hydrogen-like impurities, with the increase of their concentration, the quantity $\varepsilon_3$ in Eq. \eqref{eq:01} decreases after reaching a maximum value (see, e.g., Ref.~\cite{Shklovskii84}). The model~\cite{Poklonskij00FTT432} (see also outlines~\cite{Abboudy95}) allows one to describe such a decrease at the expense of the broadening $\delta E_\tau$ of the energy levels of single acceptors with the increase of their concentration due to the finite time of localization of hole on the acceptor. In this case, the realization of the acceptor level resonance condition \eqref{eq:16} becomes more likely, because $\delta E_\tau$ defined by Eq. \eqref{eq:18} becomes comparable with acceptor band width $W$ given by Eq. \eqref{eq:06}.

\begin{figure}
\noindent\hfil\includegraphics{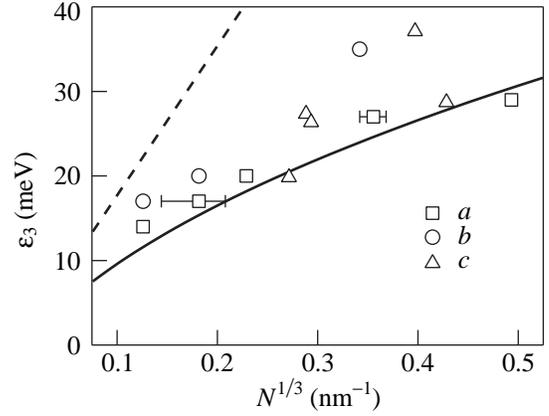} 
\caption{Solid line is the dependence of the activation energy $\varepsilon_3$ of hopping conductivity on the concentration $N$ of boron atoms in diamond, calculated by Eq. \eqref{eq:23} for the compensation ratio $K = 0.25$ at temperature $T_\text{h} = T_\text{j}/2$ defined by Eq. \eqref{eq:24}. Experimental data of papers~\cite{Borst96}, \cite{Malta95}, \cite{Visser92} are denoted by $a$, $b$, $c$, respectively. Dashed line is the calculation using the model~\cite{Shklovskii84}}\label{fig:02}
\end{figure}

\section{Conclusions}

We have developed a model of dc hopping conductivity of holes via acceptors (boron atoms) in diamond crystals. It was supposed that boron atoms with concentration $N$ and donors with concentration $KN$ form a common nonstoichiometric simple cubic lattice within the crystalline matrix. The translational period of impurity lattice is $R_\text{h} = [(1 + K)N]^{-1/3}$. In this case, only hole hops of length $R_\text{h}$ between acceptors at the instances of accidental alignment of their energy levels due to the thermal fluctuations were taken into account. The hopping conductivity was averaged over all possible orientations of the cubic impurity lattice with respect to the direction of the external electric field. It was taken into account that the energy levels of boron atoms are distributed due to the Coulomb interaction between the ions in the first coordination sphere of the impurity lattice. We assumed that the acceptor band width is much larger than the quantum broadening of acceptor energy levels because of the finite time of hole localization on boron atoms. Our calculations of the pre-exponential factor $\sigma_3$ and the activation energy $\varepsilon_3$ of the hole hopping transport $\sigma_\text{h}$ over boron atoms depending on their concentration agree well with experimental data~\cite{Borst96, Malta95, Visser92} for moderately compensated ($K \approx 0.25$) diamond crystals. The calculation was performed for temperature $T_\text{h} = T_\text{j}/2$, two times lower than the transition temperature $T_\text{j}$ from $v$-band dc hole conduction regime to hopping over boron atoms.

\section*{Acknowledgments}
The work was supported by the Belarusian program ``Nano\-tech'' and by the Russian Federation  Presidential Foundation (grant 2951.2008.2).

\section*{Appendix}

To obtain formula \eqref{eq:16}, let us supplement the notations of the main text of the paper. The probability of finding an acceptor in charge state ($-1$) is equal to $P\{-1\} = K$, and the probability of finding an acceptor in charge state ($0$) is equal to $P\{0\} = 1 - K$. The conditional probability that an acceptor with energy level $s = u$ is in charge state ($0$) is equal to $P\{0\,|\,s = u\} = f_0(u)$. The conditional probability that an acceptor with energy level $s = u$ is in charge state ($-1$) is equal to $P\{-1\,|\,s = u\} = f_{-1}(u)$.

The conditional probability of an event $A$ occurring, provided that event $B$ has occurred, has a form~\cite{Koks66, Uittl82}: $P\{A|B\} = P\{A \cap B\}/P\{B\}$. Hence, exchanging $A$ and $B$, we have $P\{A \cap B\} = P\{B|A\}P\{A\}$, and consequently
\begin{equation}
	P\{A|B\} = \frac{P\{B|A\}P\{A\}}{P\{B\}}. \tag{A.1}
\end{equation}

Formula \eqref{eq:16} has a form (A.1), where the event $B$ corresponds to one acceptor in an acceptor pair being in charge state ($0$) and the other being in charge state ($-1$); the event $A$ corresponds to each acceptor in the pair having the energy level $u$. Thus,
\begin{equation}
	P\{B\} = P\{\text{pair}(0,-1)\} = 2K(1 - K)  \tag{A.2}
\end{equation}
and the conditional probability 
\begin{align}
	P\{B|A\} = P\{\text{pair}(0,-1)\,|\,s = u\}& \notag\\
	= 2P\{0\,|\,s = u\}P\{-1\,|\,s = u\}& = 2f_0(u)f_{-1}(u). \tag{A.3}
\end{align}

Because the distribution of acceptor energy levels relative to their average value ($s = 0$) is continuous, instead of the probability $P\{A\}$ we use the probability of finding the acceptor energy level $s$ within the interval $(u, u + \mathrm{d}u)$: 
\begin{equation}
	P\{s \in (u, u + \mathrm{d}u)\} = \mathcal{G}(u)\,\mathrm{d}u.   \tag{A.4}
\end{equation}

Now the conditional probability to find a pair of acceptors in charge states ($0$) and ($-1$) with energy levels within the interval $(u, u + \mathrm{d}u)$ can be obtained by substitution of Eqs. (A.2)--(A.4) into Eq. (A.1):
\[
	P\{s \in (u, u + \mathrm{d}u)\,|\,\text{pair}(0, -1)\} = \frac{2\mathcal{G}(u)f_0(u)f_{-1}(u)\,\mathrm{d}u}{2K(1 - K)}.
\]

Thus, the quantity $\cfrac{f_0(u)f_{-1}(u)}{K(1 - K)}\,\mathcal{G}(u)\,\mathrm{d}u = P(u)\,\mathrm{d}u$ gives the conditional probability that a pair of acceptors in charge states ($0$) and ($-1$) have their energy levels within the interval $(u, u + \mathrm{d}u)$,  i.e., we obtain formula \eqref{eq:16}.



\end{document}